\documentstyle[prb,epsf,aps,multicol]{revtex}
\draft

\def\eeq{\end{equation}}

\def\prb{Phys. Rev. {\bf B}}

\def\pla{Phys. Lett. A } 
\def\pb{Physica B}
  
\def\mpl{Mod. Phys. Lett. {\bf B}}

\def\pjp{Pramana J. Phys.}
\begin{document}

\draft

\title{Quantum current enhancement effect in hybrid rings at
  equilibrium}

\author{Colin Benjamin\cite{coline} and A. M. Jayannavar\cite{amje} }

\address{Institute of Physics, Sachivalaya Marg, Bhubaneswar 751 005,
  Orissa, India}

\date{\today}
  
\maketitle
 
\begin{abstract}
  Current enhancement- a novel quantum phenomena is found to occur in
  a mesoscopic hybrid ring at equilibrium. The hybrid system is
  described by a ring with bubble which is in turn coupled to a
  reservoir. In the system the ring encloses a magnetic flux $\Phi$
  while the bubble does not enclose any flux. The novelty of this work
  lies in the fact that while earlier current enhancement was observed
  in non-equilibrium systems (e.g., a ring coupled to two reservoirs
  at different chemical potentials $\mu_{1}$ and $\mu_{2}$), herein we
  prove that current enhancement can also arise in equilibrium. In
  addition, we show that the closed system analog of our chosen open
  hybrid ring system violates parity effects. Finally, we bring to
  focus the discrepancy between the equilibrium magnetic moment
  (obtained via energy eigenvalues) and that calculated from the
  currents in the system.
\end{abstract}
\begin{multicols}{2}
  The physics of low dimensional systems particularly those whose
  system size is less than the electron phase coherence length has
  been quite vibrant in recent years thanks to technological advances
  in the field of
  nanoscience\cite{imry,datta,dephase:webb_ap,deo,bil}. The study of
  such systems where the electron retains its wave nature over the
  entire sample is termed mesoscopic physics. In these systems
  experiments have revealed that several classical laws which hold for
  macroscopic systems breakdown\cite{datta}. This is attributed to the
  interference effects of electronic waves. One of the simple quantum
  mechanical phenomena which has been predicted in such systems is
  that of current enhancement or magnification
  \cite{deo_cm,pareek_cm,colin_cm}. Current enhancement can be defined
  as follows- In a metallic loop(see inset fig.~1) connected to two
  ideal leads transport current I flows through the system.  Currents
  $I_{1}$ and $I_{2}$ flow in the upper and lower arms of the ring
  respectively. In general, $I_{1}$ is not equal to $I_{2}$ but
  $I=I_{1}+I_{2}$, Kirchoff's law. In classical case both $I_{1}$ and
  $ I_{2}$ are positive and flow in same direction as the applied
  bias. In quantum mechanics, for particular values of Fermi energy
  $I_{1}$ or $I_{2}$ can become much larger than I, this implies to
  obey Kirchoff's law the current in the other arm must be negative.
  The property that current in one of the arms is larger than the
  transport current is referred to as current enhancement effect. In
  this situation, we interpret the negative current flowing in one arm
  of the ring as a circulating current that flows continually in the
  loop.  When the negative current flows in the upper arm the
  circulating current direction is taken to be anti-clockwise (or
  negative) and when it flows in the lower arm the circulating current
  direction is taken to be clockwise (or
  positive)\cite{physicapareek}.

  The current enhancement effect leads to enhanced magnetic response
  (orbital magnetic moment) of a loop carrying current in the absence
  of magnetic flux which can lead to an experimental verification of
  this\cite{deo}. It is to be noted that these circulating currents
  arise in the absence of magnetic flux and in presence of transport
  currents (i.e., in a non-equilibrium system).  In the present work
  our thrust is whether we can observe the aforesaid current
  enhancement effect and the resulting circulating currents in
  equilibrium. For this we consider the one dimensional hybrid ring
  system as depicted in figure~1 connected to a reservoir at chemical
  potential $\mu$.  The static localised flux piercing the loop is
  necessary to break the time reversal symmetry and induce a
  persistent current in the system. The reservoir acts as an inelastic
  scatterer and as a source of energy dissipation. All the scattering
  processes in the leads including the loop are assumed to be elastic.
  The loops J1J2aJ3J1 and J1J2bJ3J1 enclose the localised flux $\Phi$.
  However, the bubble J2aJ3bJ2 does not enclose the flux $\Phi$. This
  special situation we have considered, so as to answer the question
  of existence of circulating currents in equilibrium systems.  We
  show that circulating currents (due to current enhancement) arise in
  a bubble which does not enclose a magnetic flux. We would like to
  mention here that the current enhancement effect and the associated
  circulating currents arise even when the magnetic field extends over
  the entire sample. However, for this the treatment is involved as
  one has to study separately persistent as well as circulating
  currents in the bubble as they have different symmetry properties.
  This has been studied in a simple loop in the presence of both
  transport currents and magnetic flux\cite{physicapareek}.

  In the local coordinate system the wave-functions in the various
  regions of the ring in absence of magnetic flux are given as follows

\begin{eqnarray}
\psi_0&=&  e^{ikx_{0}}+r e^{-ikx_{0}},\nonumber\\
\psi_j&=&a_{j} e^{i(k+\frac{\alpha_{j}}{l_{j}})x_{j}}+b_{j}
e^{-ikx_{j}+i\frac{\alpha_{j}}{l_{j}}(x_{j}-l_{j})}.
\end{eqnarray} 

Here $x_{j},j=1,..4$ are coordinates along the the segments J1J2,
J2bJ3, J2aJ4 and J3J1 respectively and $x_{0}$ is the coordinate along
the connecting lead to the reservoir, while $\alpha_{j}$'s are the
phases picked up by the electron as it traverses the various regions
of the system with the restriction that
$\alpha_{1}+\alpha_{2}+\alpha_{4}=2\pi\Phi/\Phi_{0},$ and
$\alpha_{1}+\alpha_{3}+\alpha_{4}=2\pi\Phi/\Phi_{0}$ which implies
$\alpha_{2}=\alpha_{3}$. To solve for the unknown coefficients in
eqn.(1) we use Griffith\cite{grif} boundary condition at the junctions
$J1, J2$ and $J3$. These boundary conditions are due to the continuity
of wavefunctions and conservation of current (Kirchoff's
law)\cite{xia}.

In the lead connecting the reservoir to our circuit there is no
current flow as $|r|^2=1$.  The current densities (dimensionless
form)\cite{physicapareek,buti} in the small interval $dk$ around the
Fermi energy $k$ in the various segments of the circuit are given by -
$I_{j}=|a_{j}|^2-|b_{j}|^2,$ The current densities are calculated from
the usual formula of current density in presence of magnetic flux-
$J_{j}=\frac{e\hbar}{2mi}(\psi_{j}^*\nabla\psi_{j}-\psi_{j}\nabla\psi_{j}^*
-{2i}\frac{\alpha_{j}}{l_{j}}\psi_{j}^*\psi_{j})$, which implies
$I_{j}=\frac{J_{j}}{e\hbar k/m}$.

The persistent current densities in various parts of the system show
cyclic variation with flux and $\Phi_{0}$ periodicity (reminiscent of
Aharonov-Bohm oscillations), and oscillate between positive and
negative values as a function of energy or the wave-vector $k$ as
expected. Since the analytical expressions for these currents are too
lengthy we confine ourselves to a graphical interpretation of the
results. It should be noted that in all these expressions for current
densities, flux enters only through the combinations
$\alpha_{1}+\alpha_{2}+\alpha_{4}$ and
$\alpha_{1}+\alpha_{3}+\alpha_{4}$, the magnitude of these
combinations is given by $2\pi\Phi/\Phi_{0}$ as expected. For us the
current densities in the bubble, i.e., J2bJ3aJ2 are of special
importance as in this region there is possibility of current
enhancement which would be analysed below. The current density shows
an extremum near the corresponding eigen-states of the system. We have
calculated these eigen states for two different cases. For open system
as depicted in figure~1, one can calculate the energies (or
wave-vector) of these states by looking at the poles of the S-Matrix.
In our case S-Matrix is simply a complex reflection amplitude $r$. We
have also analysed the eigen states of a closed system (without
coupling lead to reservoir).

We analyse the case of a bubble with unequal lengths, of its two arms
i.e the length of $J2bJ3 \neq J2aJ3$. This asymmetry implies that
currents in the two arms of the bubble are not equal. In figure~2 we
plot the persistent current densities in various parts of the circuit.
It should be noted that absolute value of the persistent current
densities $I_{2}$ and $I_{3}$ are individually much larger than the
input current density $I_{1}$ into the bubble and thus the current
enhancement effect is evident (without violating the basic Kirchoff's
law). The input current arises due to the presence of flux $\Phi$ as
it breaks the time reversal symmetry. The system parameters are
mentioned in the figure caption.  In the interval, $5.2<kL<7.4$
current density $I_{1}$ changes from positive to negative and exhibits
extremum around the real part of the poles of the S-Matrix.  When
$I_{1}$ is positive, negative current density of magnitude $I_{2}$
flows in the arm $J2bJ3$ of the bubble. Thus, when $I_{1}$ is positive
circulating current flows in the anti-clockwise direction in the
bubble. In the range wherein $I_{1}$ is negative, i.e, input current
into the bubble is in anti-clockwise direction, then positive current
flows in arm $J2aJ3$. According, to our convention as mentioned
earlier, circulating current flows in the anti-clockwise direction.
In all the figures drawn, the length of the bubble is $l=l_{2}+l_{3}$
which is taken as unity throughout our discussion. The current
densities along with the Fermi wave-vectors are in their dimensionless
form. Of course the phenomena of current enhancement is extremely
sensitive to the arm lengths of the bubble. It should be noted that if
we interchange the values of $l_{2}$ and $l_{3}$ keeping other
parameters unchanged circulating current will flow in a clockwise
direction. This is obvious from the geometry of the problem.
Alongwith the current densities the persistent currents in various
parts of the ring can also be plotted, to do that we integrate the
current densities $J_{j}$ in various regions of the circuit over the
Fermi wave vector $k_{f}$.  The persistent currents $P_{j}$ at
temperature $T=0$ is given by

\begin{eqnarray}
P_{j}=-\int_{0}^{k_{f}}dk J_{j} 
\end{eqnarray}

In figure~3 we have plotted the persistent currents(in dimensionless
units) for the system parameters as mentioned in the caption. An
interesting point to note is that although in most cases current
enhancement occurs around the eigenenergies of the closed system there
are a few exceptions. In figure~4 we plot one of those exceptions.
Herein, we show that current enhancement does not occur at a place
which is an eigen'k' of the aforesaid system. Here the eigen
wave-vector $kL$ corresponds to $13.85$. One can readily notice that
the persistent current density (i.e, input current density $I_{1}$
into the bubble) shows extrema around this value. In this region both
the currents in the bubble $I_{2}$ and $I_{3}$ are individually
smaller than $I_{1}$ and they flow in the same direction as the input
current.  Hence we do not observe current enhancement.

The above discussion shows that current enhancement effect is not
restricted to non-equilibrium systems only but is also expressed in
mesoscopic systems at equilibrium. Our special system not only shows
current enhancement but also in it parity effects are violated, of
course in its closed system disguise. The study of parity effects or
rather the lack of it especially in a metallic ring with a static
localised flux at its center and a stub attached to it has attracted a
lot of interest\cite{deo_bpe}. Parity effects are defined as follows-
The persistent current carried by an electron in the eigen state
$E_{n}$ in case of an isolated single loop is defined as
$I_{n}=-\frac{1}{c}\frac{\partial E_{n}}{\partial\Phi}$. In a closed
single loop persistent current changes its sign as we go from one
level to the next successive level, i.e., from diamagnetic to
paramagnetic or vice-versa.  The total persistent current is given by
sum of currents carried by all electrons. Thus for spinless electrons
at temperature ($T=0$), a system with odd number of electrons behaves
as a diamagnet while that with an even number of electrons behaves as
a paramagnet. This is called parity effect.  In figure~5 we have
plotted the first few eigen energies $E=k_{n}^2$ of our isolated
system (with the connecting lead to reservoir removed).  The system
parameters are mentioned in the figure caption. These eigen energies
are calculated from the condition that the determinant of the
coefficient matrix must vanish. The coefficient matrix is built from
first principles using quantum waveguide theory with the second
wavefunction of equation~1. The eigen energies are flux periodic with
period $\Phi_{0}$.  For our system we immediately notice that certain
number of successive energy levels from the bottom, have same
direction of slope with respect to flux. Particularly the third and
fourth energy levels from bottom carry diamagnetic current for small
values of flux while levels five and six again from bottom carry a
paramagnetic current. Thus breaking the well known parity effect. For
details we refer to Ref.[\onlinecite{deo_bpe}].

As a corollary to the discussion above, we consider the magnetic
moments of our system. The orbital magnetic moment defined via
currents in a loop, depends strongly on the topology of the system. If
we consider our system as depicted in figure~1 to be planar and lying
in the x-y plane then the magnetic moment density would be
$\mu=\frac{1}{c}(I_{1}A_{r}+I_{3}A_{b})$ wherein $A_{r}$ and $A_{b}$
are areas enclosed by the ring($J1J2aJ3J1$) and bubble($J2aJ3bJ2$)
respectively, for details we refer the interested reader to Ref.
[\onlinecite{cedraschi}]. Another orientation of the system in which
the ring is not to the left but to the right of the bubble gives
$\mu=\frac{1}{c}(-I_{1}A_{r}^\prime-I_{2}A_{b})$, where $A_{r}^\prime$
is the area of the ring for this changed configuration. Several other
orientations also are possible, for example, if the bubble lies in x-y
plane and the ring lies in x-z plane, then the magnetic moment density
for the bubble is $\mu_{z}=\frac{1}{c}(I_{3}A_{b}-I_{2}A_{b})/2$ and
in case of the ring is $\mu_{y}=\frac{1}{c}I_{1}A_{r}$ . All the above
examples buttress the fact that the orbital magnetic moment is
inherently linked to the topology of the system. An important point to
be noted is that the magnetic moment calculated herein is not same
(qualitatively), as the magnetic moment calculated in case of the
closed system from its eigenenergy spectra, which is same for all
topological situations. A full discussion on this discrepancy
alongwith a detailed analysis of different quantum effects exhibited
by our chosen mesoscopic system will be presented elsewhere.

In conclusion, we have shown that the phenomenon of current
enhancement is not restricted to non-equilibrium mesoscopic systems
only but can also arise in equilibrium systems but ofcourse in the
presence of magnetic flux. In addition to this quantum effect our
hybrid ring geometry breaks parity effects in its closed system
analog.

\begin{figure}
  \protect\centerline{\epsfxsize=3.5in \epsfbox{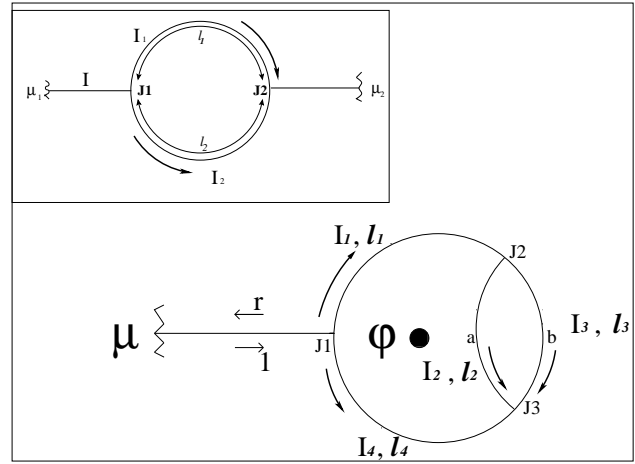}}
  \vskip 1.0in
\caption{The hybrid ring system connected to a reservoir at chemical 
  potential $\mu$. The bubble is denoted by the structure J2bJ3aJ2.
  The localised flux $\Phi$ penetrates the ring. The current densities
  in various parts of the structure are denoted by I's while the
  lengths of the various regions are denoted by l's. In the inset we
  have shown the non-equilibrium case, a one dimensional mesoscopic
  ring with leads is connected to two reservoirs at chemical
  potentials $\mu_{1}$ and $\mu_{2}$.}

\end{figure}

\begin{figure}
  \protect\centerline{\epsfxsize=3.5in \epsfbox{ 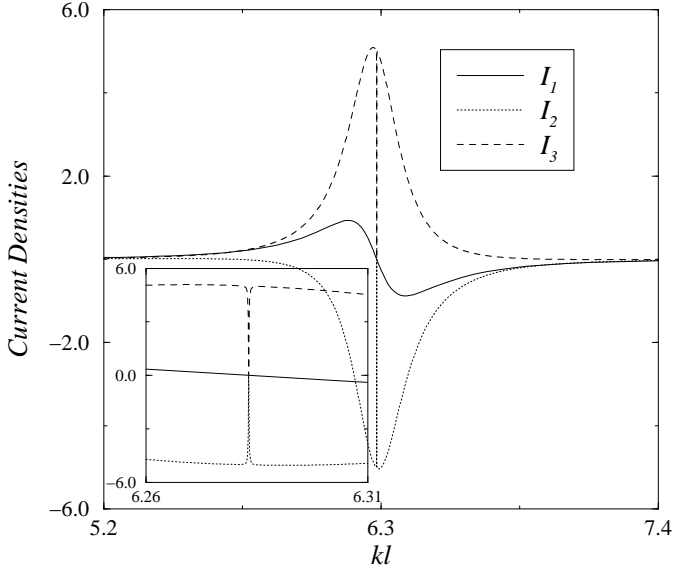}}
\caption{Current enhancement shown with lengths
  $l_{1}/l=l_{4}/l=0.75, l_{2}/l=0.45, l_{3}/l=0.55$. Herein the
  persistent current densities in the various parts of the circuit are
  plotted as function of the dimensionless Fermi wavevector $kl$. The
  persistent current density in $J1J2$ is denoted by the solid line
  while those in $J2bJ3$ and $J2aJ3$ are denoted by dotted and dashed
  line. Flux $\Phi=0.1$.}
\end{figure}

\begin{figure}
  \protect\centerline{\epsfxsize=3.5in \epsfbox{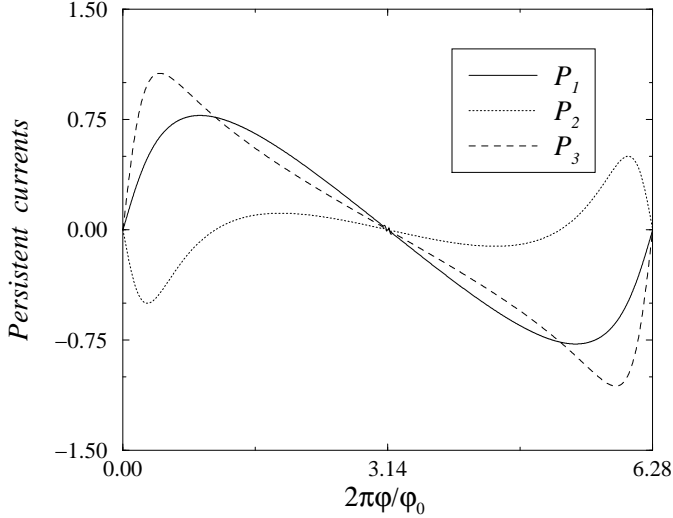}}
\caption{Current enhancement shown with lengths 
  $l_{1}/l=l_{4}/l=0.25, l_{2}/l=0.45, l_{3}/l=0.55$. Herein the
  persistent currents in the various parts of the circuit are plotted
  as function of flux. The Fermi wavevector here is $k_{f}=2\pi$.}
\end{figure}

\begin{figure}
  \protect\centerline{\epsfxsize=3.5in \epsfbox{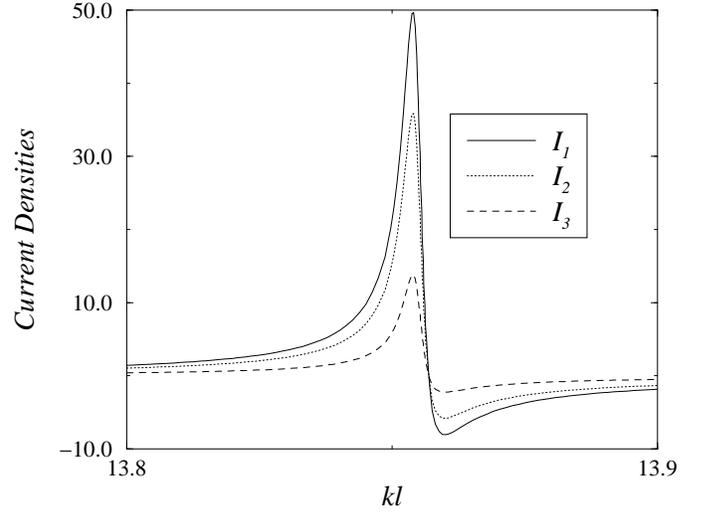}}
\caption{Absence of  current enhancement shown with lengths
  $l_{1}/l=l_{4}/l=0.75, l_{2}/l=0.25, l_{3}/l=0.75$. Herein the
  persistent current densities in the various parts of the circuit are
  plotted. The persistent current density in $J1J2$ is denoted by the
  solid line while those in $J2aJ3$ and $J2bJ3$ are denoted by dotted
  and dashed line. The $kl$ value $13.85$ is an eigen wavevector of
  the closed system. Flux $\Phi=0.1$.}
\end{figure}

\begin{figure}
\protect\centerline{\epsfxsize=3.5in \epsfbox{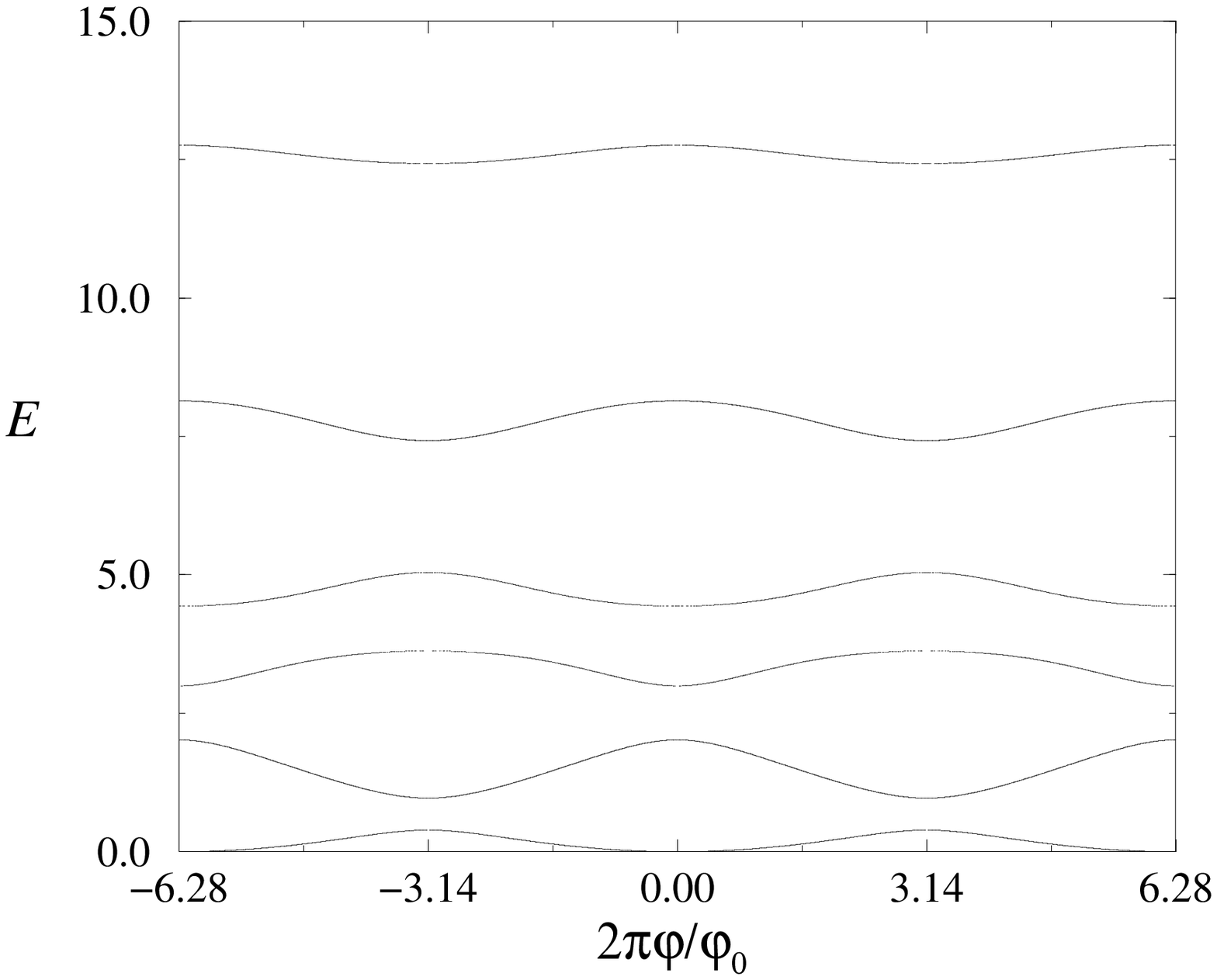}}
\caption{Breakdown of parity effects in the closed form of our open
  hybrid ring system as depicted in figure~1. The length parameters
  are $l_{1}/l=0.75, l_{2}/l=0.35, l_{3}/l=0.65$. The energies are
  normalised by $\pi^{2}$.}

\end{figure}

\end{multicols}


\begin{references}
  
\bibitem[\dagger]{coline} Electronic address: colin@iopb.res.in
  
\bibitem[\star]{amje} Electronic address: jayan@iopb.res.in
  
\bibitem{imry} Y. Imry, {\it Introduction to Mesoscopic Physics}
  (Oxford University, New York, 1997).
  
\bibitem{datta} S. Datta, {\it Electronic transport in mesoscopic
    systems} (Cambridge University Press, Cambridge, 1995).
  
\bibitem{dephase:webb_ap} S. Washburn and R. A. Webb, {\it Adv. Phys.}
  {\bf 35}, 375.
  
\bibitem{deo} P. S. Deo and A. M. Jayannavar, \pjp {\bf 56}, 439
  (2001).
  
  
\bibitem{bil} M. B\"{u}ttiker, Y.Imry and R.Landauer , \pla{\bf 96},
  365 (1983).
  
\bibitem{deo_cm} A. M. Jayannavar and P. S. Deo, \prb {\bf 49}, 13685
  (1994).
  
\bibitem{pareek_cm} T. P. Pareek, P. S. Deo and A. M.  Jayannavar,
  \prb {\bf 52}, 14 657 (1995).
 
\bibitem{colin_cm} C. Benjamin, S. K.  Joshi, D.Sahoo and A. M.
  Jayannavar, \mpl{\bf15}, 19 (2001).
  
\bibitem{physicapareek} A. M. Jayannavar, P. Sinngha Deo, and T. P.
  Pareek, in Proceedings of International Workshop on Novel Physics in
  Low-Dimensional Electron Systems, Madras, India [\pb {\bf212}, 216
  (1995)].
  
\bibitem{grif}S. Griffith, Trans. Faraday. Soc.{\bf 49}, 650 (1953)
  
\bibitem{xia}J-B. Xia, \prb {\bf 45}, 3593 (1992).
  
\bibitem{buti} M. B\"{u}ttiker, \prb{\bf 32}, 1846 (1985).
  

\bibitem{deo_bpe} P. S. Deo, \prb{\bf 51}, 5441 (1995).

\bibitem{cedraschi} P. Cedraschi and M. B\"{u}ttiker, \prb{\bf 63},
  165312 (2001).

\end{references}
\end{document}